\documentclass[aps,prl,reprint,superscriptaddress,nofootinbib,floatfix]{revtex4-2}
\usepackage[utf8]{inputenc}

\usepackage{silence}
\usepackage{amsmath}
\usepackage{amssymb}
\usepackage{graphicx}
\usepackage{hyperref}
\usepackage{booktabs}
\usepackage{underscore}
\AtBeginDocument{%
  \heavyrulewidth=.08em
  \lightrulewidth=.05em
  \cmidrulewidth=.03em
  \belowrulesep=.65ex
  \belowbottomsep=0pt
  \aboverulesep=.4ex
  \abovetopsep=0pt
  \cmidrulesep=\doublerulesep
  \cmidrulekern=.5em
  \defaultaddspace=.5em
}
\usepackage{csquotes}
\usepackage{microtype}
\usepackage[dvipsnames, table]{xcolor}
\usepackage[normalem]{ulem}

\newcommand{\lcdm}{\ensuremath{\Lambda\mathrm{CDM}}}
\newcommand{\mb}{\ensuremath{M_B}}
\newcommand{\kmsmpc}{\ensuremath{\mathrm{km\,s^{-1}\,Mpc^{-1}}}}

\begin{document}
\begin{flushright}
        {\large \tt TTK-26-22} \qquad \qquad 
        {\large \tt TTP26-27}
\end{flushright}
\vspace*{-1cm}

\title{The \texorpdfstring{$H_0$}{H0} World Cup. I. Summary of the baseline group stage results}

\author{Nils Sch\"oneberg}
\affiliation{University Observatory, Faculty of Physics, Ludwig-Maximilians-Universit\"at,
Scheinerstr. 1, 81677 M\"unchen, Germany}
\affiliation{Excellence Cluster ORIGINS, Boltzmannstrasse 2, 85748 Garching, Germany}
\author{Vivian Poulin}
\affiliation{Laboratoire Univers et Particules de Montpellier (LUPM), Universit\'e de
Montpellier \& CNRS, Place Eug\`ene Bataillon, 34095 Montpellier Cedex 05, France}
\author{Angelo G. Ferrari}
\affiliation{INFN--Bologna, Via C. Berti Pichat 6/2, 40127 Bologna, Italy}
\author{Fabio Finelli}
\affiliation{INAF--Osservatorio di Astrofisica e Scienza dello Spazio di Bologna,
Via Piero Gobetti 101, 40129 Bologna, Italy}
\affiliation{INFN--Bologna, Via C. Berti Pichat 6/2, 40127 Bologna, Italy}
\author{Julien Lesgourgues}
\affiliation{Institute for Theoretical Particle Physics and Cosmology (TTK), RWTH Aachen
University, D-52056 Aachen, Germany}
\author{Luca Morelli}
\affiliation{Dipartimento di Fisica e Astronomia “Augusto Righi”, Università di Bologna, Via Piero Gobetti 93/2, I-40129 Bologna, Italy}
\affiliation{INAF--Osservatorio di Astrofisica e Scienza dello Spazio di Bologna,
Via Piero Gobetti 101, 40129 Bologna, Italy}
\affiliation{INFN--Bologna, Via C. Berti Pichat 6/2, 40127 Bologna, Italy}
\author{Markus R. Mosbech}
\affiliation{Institute for Theoretical Particle Physics and Cosmology (TTK), RWTH Aachen
University, D-52056 Aachen, Germany}
\affiliation{Institute for Theoretical Particle Physics (TTP), Karlsruhe Institute of
Technology (KIT), 76128 Karlsruhe, Germany}
\author{Ravi Kumar Sharma}
\affiliation{Institute for Theoretical Particle Physics and Cosmology (TTK), RWTH Aachen
University, D-52056 Aachen, Germany}
\author{Th\'eo Simon}
\affiliation{Laboratoire Univers et Particules de Montpellier (LUPM), Universit\'e de
Montpellier \& CNRS, Place Eug\`ene Bataillon, 34095 Montpellier Cedex 05, France}
\affiliation{Laboratoire de Physique Nucléaire et de Hautes Energies (LPNHE), CNRS/IN2P3 \& Sorbonne Université, 4 place Jussieu, 75005 Paris, France}

\begin{abstract}
The Hubble tension has reached a nominal significance above $7\sigma$, while new
high-precision measurements of the cosmic microwave background (CMB) and baryon
acoustic oscillations (BAO) sharpen the test of proposed solutions. Using a common framework, we compare
fourteen representative alternatives to the standard $\Lambda$ Cold Dark Matter (\lcdm) model in light of up-to-date CMB, BAO and supernovae data to gauge their ability to resolve the tension. 
The models span late-time modifications, modified recombination, and exotic
pre-recombination expansion histories driven by additional radiation or a localized dark energy injection.
We evaluate each proposal with complementary frequentist and Bayesian measures of
the residual calibration tension and of the improvement in the joint fit. 
Both
approaches identify the same broad hierarchy. 
Early dark energy and early modified gravity
models perform best, shifting the $H_0$ inference without local measurement priors toward
$70\,\kmsmpc$ and reducing the residual discrepancy with SH0ES to approximately
$2.5$--$3.6\sigma$, depending on the model and statistic, while receiving strong
support over \lcdm{} in the combined fit. Varying the electron mass at
recombination yields an intermediate improvement, whereas the enhanced-radiation and late-time scenarios do not improve over \lcdm{}. 
This Letter summarizes the {\it group stage} of the
competition; in a companion paper (Paper II)~\cite{H0WorldCup:companion} we present the results of an exhaustive set of analyses and assess their robustness to variations in modeling assumptions and datasets.
\end{abstract}

\maketitle

The disagreement between the locally calibrated expansion rate and the value
inferred from early-Universe observations within \lcdm{} remains one of the most
persistent problems in cosmology. A recent joint analysis of local methods gives
$H_0=73.50\pm0.81\,\kmsmpc$ \cite{H0DN:2025lyy}, whereas the combination of
current CMB measurements interpreted in \lcdm{} yields
$H_0=67.19\pm0.38\,\kmsmpc$ \cite{SPT-3G:2025bzu}. Taken at face value, the
discrepancy has reached $7.1\sigma$. The proliferation of proposed
solutions~\cite{DiValentino:2021izs,CosmoVerseNetwork:2025alb}, each typically
confronted with different datasets and analysis choices, makes a controlled
comparison essential: a successful model must both reduce the tension and
provide a sufficiently improved description of the joint data to justify its
additional parameters. Such a comparison is only meaningful
when every contender faces the same likelihoods in the same analysis pipeline.

We update the $H_0$ Olympics \cite{Schoneberg:2021qvd}, an attempt at benchmarking model performance (see also Ref.~\cite{Escudero:2022rbq,Khalife:2023qbu,AtacamaCosmologyTelescope:2025nti}), using current CMB data \cite{Planck:2020olo,ACT:2025blo,SPT-3G:2025bzu,ACT:2023kun,SPT-3G:2024atg},
BAO data from the Dark Energy Spectroscopic Instrument (DESI) \cite{DESI:2025zgx,DESI:2025zpo}, and Pantheon+ Supernovae (SN) data \cite{Brout:2022vxf}. We refer to this comparison as the
\enquote{$H_0$ World Cup}, presented in two parts. This Letter (Paper I) is a
condensed summary of the {\it group stage}: each contender is 
compared to CMB+BAO+SN data, with and without inclusion of local $H_0$ measurements modeled as a prior on the SN magnitude \mb{}. We examine whether the extended models reduce the tension and are preferred over \lcdm{} in the combined analysis from both a frequentist and Bayesian statistical perspective.
In the companion Paper II \cite{H0WorldCup:companion}, we present
the full model descriptions, priors, and analysis details, and we study
additional extensions of the contenders. We study in particular the combination of
early- and late-time solutions through spatial curvature or an evolving dark
energy equation of state, together with the interplay with the DESI
preference for dynamical dark energy \cite{DESI:2025fii} and with
cosmological neutrino-mass bounds. Paper II also establishes the robustness
of the group-stage ranking by subjecting the qualifiers to a knockout round of
dataset variations---extended Planck multipole ranges, removal of Atacama Cosmology telescope (ACT) data,
alternative supernova calibrations, and weak-lensing ($S_8$) and big bang
nucleosynthesis constraints.

The fourteen contenders represent five physical strategies. Group E modifies
the pre-recombination expansion rate through a transient, non-radiative energy
component: axion-like early dark energy
(EDE)~\cite{Doran:2000jt,Wetterich:2004pv,Doran:2006kp,Kamionkowski:2014zda,
Karwal:2016vyq,Poulin:2018dzj,Poulin:2018cxd,Smith:2019ihp}, whose many
realizations and
extensions~\cite{Agrawal:2019lmo,Lin:2019qug,Alexander:2019rsc,
Sakstein:2019fmf,Das:2020wfe,Niedermann:2019olb,Niedermann:2020dwg,
Niedermann:2021vgd,Ye:2020btb,Simon:2023hlp,Berghaus:2019cls,Freese:2021rjq,
Braglia:2020bym,Sabla:2021nfy,Sabla:2022xzj,Gomez-Valent:2021cbe,
Moss:2021obd,Guendelman:2022cop,Karwal:2021vpk,McDonough:2021pdg,
Wang:2022nap,Alexander:2022own,McDonough:2022pku,Nakagawa:2022knn,
Gomez-Valent:2022bku,MohseniSadjadi:2022pfz,Kojima:2022fgo,Rudelius:2022gyu,
Oikonomou:2020qah,Tian:2021omz,Maziashvili:2021mbm} and confrontations with
data~\cite{Hill:2020osr,Ivanov:2020ril,DAmico:2020ods,Murgia:2020ryi,
Smith:2020rxx,Simon:2022adh,ACT:2025tim,Poulin:2025nfb,chaussidon:2025,
SPT-3G:2025vyw} are reviewed
in~\cite{Kamionkowski:2022pkx,Poulin:2023lkg,McDonough:2023qcu}; cold new
early dark energy
(NEDE)~\cite{Niedermann:2019olb,Niedermann:2020dwg,Niedermann:2023ssr,
Cruz:2023lmn,Chatrchyan:2024xjj}; early modified gravity
(EMG)~\cite{Braglia:2020bym} as a representative model respecting Solar-System constraints within scalar-tensor theory of gravity \cite{Umilta:2015cta,Ballardini:2016cvy,Rossi:2019lgt,Braglia:2020iik,
Zumalacarregui:2020cjh,Ferrari:2023qnh};
and its minimally coupled Rock 'n' Roll (RnR)
limit~\cite{Agrawal:2019lmo}. Group L contains post-recombination or
late-time possibilities: a sign-switching cosmological
constant~\cite{Akarsu:2021fol,Ibarra-Uriondo:2026zbp,
Bouhmadi-Lopez:2026ckz,Khandelwal:2026btl}, an analysis agnostic about
low-multipole reionization information~\cite{Pagano:2019tci}, and interacting
dark matter--dark
energy~\cite{Gavela:2009cy,DiValentino:2017iww,DiValentino:2017oaw,
DiValentino:2019ffd,DiValentino:2019jae}. Group E+L includes {\em thawing gravity} as a fundamental physics model modifying early and late dynamics \cite{Ye:2024zpk,Ye:2024ywg}.
Group M changes recombination through either a
varying electron mass~\cite{Hart:2019dxi,Sekiguchi:2020teg,Hart:2021kad,Seto:2022xgx,
Chluba:2023xqj,Seto:2024cgo,Baryakhtar:2024rky,Schoneberg:2024ynd,
Baryakhtar:2025uxs,Smith:2025arq,Smith:2025grk,Garramone:2026evc} or a
four-parameter deformation of the ionization
history~\cite{Mirpoorian:2024fka}, motivated by baryon clumping in the
presence of primordial magnetic
fields~\cite{Jedamzik:2020zmd,Jedamzik:2020krr,Galli:2021mxk}. Group R adds or produces radiation before
recombination: free-streaming
$\Delta N_{\rm eff}$~\cite{Bernal_2016,Planck:2018jri,Gariazzo:2023hch,
Allali:2024cji,Drewes:2024wbw,Saravanan:2025cyi}, self-interacting dark
radiation (SIDR)~\cite{Lesgourgues:2015wza,Cyr-Racine:2015ihg}, Wess--Zumino
dark radiation
(WZDR)~\cite{Wess:1974,Aloni:2021eaq,Allali:2021azp,Joseph:2022jsf,
Joseph:2022yys,Schoneberg:2022grr,Aloni:2023tff,Meiers:2023gft,
Sobotka:2023bzr,Bagherian:2024obh,Zhou:2024igb,Cvetko:2025kda,
Smith:2025zsg}, and dark-radiation--matter decoupling
(DRMD)~\cite{Weinberg:1973am,Niedermann:2021ijp,Niedermann:2021vgd,
Garny:2024ums,Garny:2025kqj}. In every case, the reference
cosmology is flat \lcdm{} with a free summed neutrino mass (modeled as 3 mass-degenerate neutrinos), and each model is
analyzed with the priors adopted for its physical parameters in
Paper II~\cite{H0WorldCup:companion}, following the parent studies cited
above.

Our uncalibrated baseline combines four observational ingredients. 
The CMB dataset uses the Planck PR3 TT likelihood at $\ell \le 29$ \cite{Planck:2019nip}, the SRoll2 EE likehood at $\ell \le 29$ \cite{Pagano:2019tci}, Planck PR4 NPIPE temperature and polarization \cite{Planck:2020olo} at multipoles $30 \le  \ell < 1000$ in TT and $30 \le \ell < 600$ in TE and EE, complemented by the ground-based measurement from the Atacama Cosmology Telescope (ACT DR6) \cite{ACT:2025blo} and the South Pole Telescope (SPT-3G D1) \cite{SPT-3G:2025bzu} at higher multipoles using the nuisance-marginalized lite likelihoods. It also includes ACT DR6 + Planck \cite{ACT:2023kun} and SPT-3G \cite{SPT-3G:2024atg} lensing likelihoods.
We add DESI DR2 BAO over $0.3<z<2.3$ \cite{DESI:2025zgx,DESI:2025zpo} and uncalibrated Pantheon+ SN
luminosity distances over $0.01<z<2.3$ \cite{Brout:2022vxf}. We denote this
combination by $A\equiv\mathrm{CMB+BAO+SN}$.

\begin{figure*}[!t]
 \centering
 \includegraphics[width=0.98\textwidth]{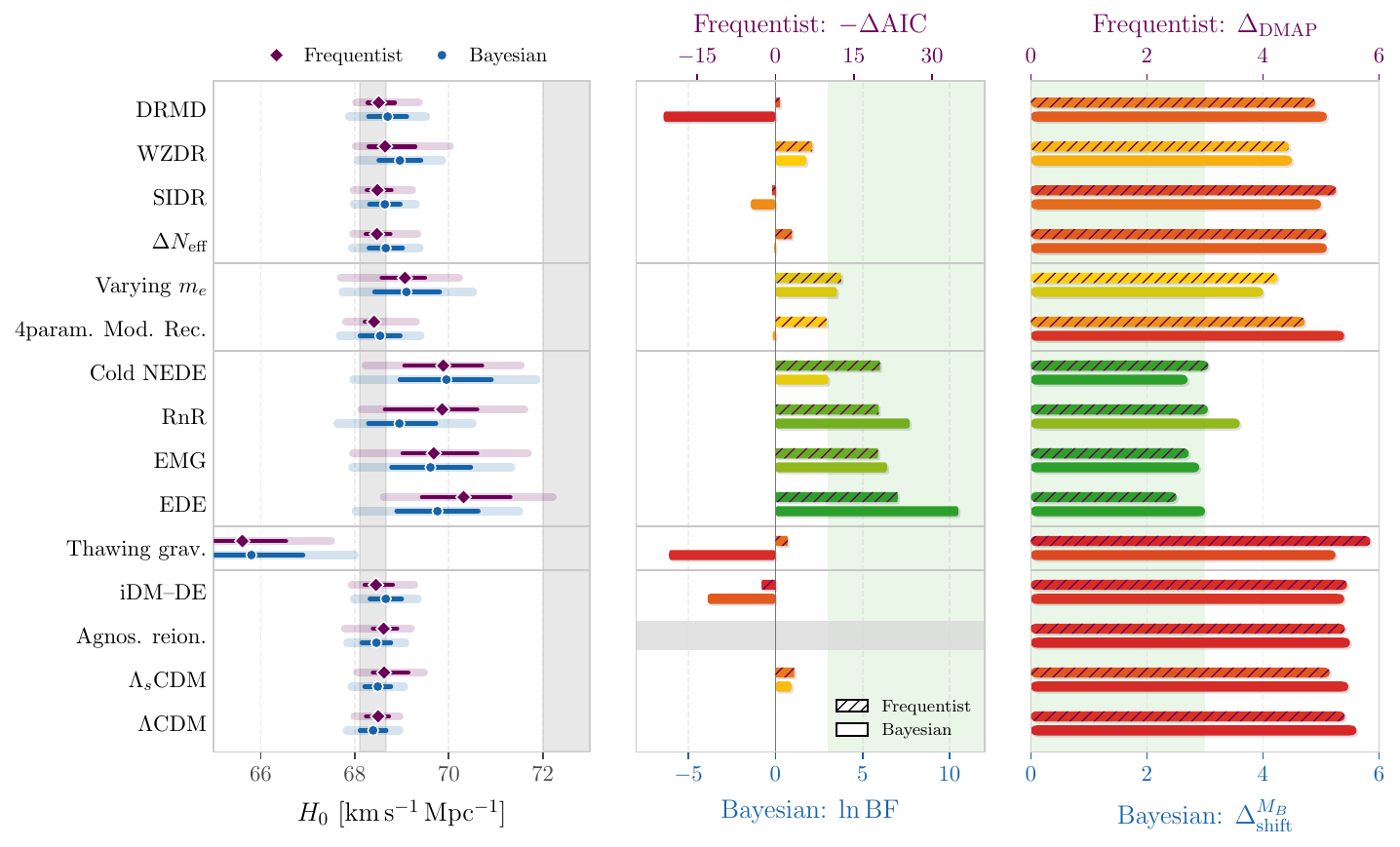}
 \caption{\label{fig:baseline-summary}Combined baseline comparison. Left:
 frequentist profile-likelihood intervals (purple diamonds) and Bayesian
 marginalized credible intervals (blue circles) for $H_0$ from CMB+BAO+SN;
 thick and thin segments denote the approximate $2\sigma$ and $1\sigma$
 ranges, respectively, and the gray bands show the \lcdm{} inference versus local measurement. Middle: model preference over $\Lambda$CDM based on CMB+BAO+SN+$\mb$ data, measured by
 $-\Delta{\rm AIC}$ (hatched bars, top scale) and $\ln{\rm BF}$ (solid bars,
 bottom scale). Right: residual tension between CMB+BAO+SN and the $\mb$
 prior, measured by $\Delta_{\rm DMAP}$ (hatched bars, top scale) and the
 Bayesian $\Delta_{\rm shift}^{M_B}$ (solid bars, bottom scale).}
\end{figure*}

On the other hand, the local calibration enters as the independent Gaussian prior $B$,
\begin{equation}
 B\equiv \mb=-19.253\pm0.027 ,
\end{equation}
taken from Ref.~\cite{Riess:2021jrx}. Applied to the full Pantheon+ sample,
this calibration gives $H_0=73.26\pm0.98\,\kmsmpc$. Keeping $A$ and
$B$ conceptually separate is important. The posterior or profile constraints on
$H_0$  shown below (see Fig.~\ref{fig:baseline-summary}) are inferred from $A$ alone; the tension and model-comparison
statistics test the joint dataset $A+B$.

We use parallel frequentist and Bayesian criteria to distinguish two questions:
1. does a model reconcile the discrepant datasets? and 2. does it improve the global fit enough
to warrant its extra freedom?  

For the frequentist analysis, we consider the non-Gaussian tension 
\begin{equation}
 \Delta_{\rm DMAP}=
 \sqrt{\chi^2_{\min,A+B}-\chi^2_{\min,A}-\chi^2_{\min,B}},
\end{equation}
and the Akaike Information Criterion (AIC) for model comparison 
\begin{equation}
 \begin{aligned}
 {\rm AIC}&=\chi^2_{\min,A+B}+2N_{\rm param},\\
 \Delta{\rm AIC}&={\rm AIC}_{\rm model}-{\rm AIC}_{\lcdm}.
 \end{aligned}
\end{equation}
A larger positive value of $-\Delta{\rm AIC}$ indicates stronger support
relative to \lcdm; $-\Delta{\rm AIC}>10$ is our benchmark for a substantial
improvement. 

For the Bayesian analysis, we consider a non-Gaussian statistic based on the posterior distribution of the parameter difference, 
\begin{equation}
 p(\Delta\theta)=\int p_A(\theta)\,p_B(\theta-\Delta\theta)\,
 {\rm d}\theta .
\end{equation}
Integrating the region with density larger than $p(\Delta\theta=0)$ and mapping
the result to an equivalent Gaussian significance defines
$\Delta_{\rm shift}$ \cite{Raveri:2021wfz}. Bayesian model comparison uses
the evidence $Z_M=\int {\cal L}(D|\theta,M)\pi(\theta|M)\,{\rm d}\theta$ and the log-Bayes factor
$\ln{\rm BF}=\ln Z_M-\ln Z_{\lcdm}$. 
Because the evidence averages over the
prior volume, its numerical value must be interpreted with the adopted priors in mind. We take $\ln{\rm BF}>3$ as our benchmark for a substantial improvement.

All analyses make use of the code \texttt{MontePython-v3}~\cite{Brinckmann:2018cvd,Audren:2012wb}, interfaced with modified versions of \texttt{CLASS} \cite{Lesgourgues:2011re}, and accelerated using the \texttt{OL\'E} emulator package~\cite{Gunther:2025xrq}. For frequentist analyses, we compute global best-fits and Profile Likelihood of $H_0$ using \texttt{procoli}\footnote{\href{https://github.com/tkarwal/procoli}{https://github.com/tkarwal/procoli}} \cite{Karwal:2024qpt} and construct credible interval using the graphical method \cite{Herold:2024ksu}. To obtain Bayesian posteriors, we run Monte Carlo Markov chains (MCMC) using \texttt{MontePython-v3}~\cite{Brinckmann:2018cvd,Audren:2012wb} and the Metropolis-Hastings algorithm, taking a Gelman-Rubin criterion $|R-1|<0.01$ for convergence~\cite{Gelman:1992zz}. We analyze chains with the package \texttt{liquidcosmo}.\footnote{\href{https://github.com/schoeneberg/liquidcosmo}{https://github.com/schoeneberg/liquidcosmo}} We computed $\Delta_\mathrm{shift}$ with \texttt{tensiometer}\footnote{\href{https://github.com/mraveri/tensiometer}{https://github.com/mraveri/tensiometer}} \cite{Raveri:2019,Raveri:2021wfz},
and $\ln{\rm BF}$ using \texttt{MCevidence}\footnote{\href{https://github.com/yabebalFantaye/MCEvidence}{https://github.com/yabebalFantaye/MCEvidence}} \cite{Heavens:2017afc}.

 Figure~\ref{fig:baseline-summary} and Table~\ref{tab:selection} summarize the frequentist
and Bayesian comparisons. Within \lcdm{}, the residual calibration tension for
this dataset combination is $\Delta_{\rm DMAP}=5.4\sigma$ and
$\Delta_{\rm shift}=5.6\sigma$, setting the reference that every contender must
improve upon. The $H_0$ intervals inferred from CMB+BAO+SN alone already
reveal a clear ordering: All models in group E (dark energy injection) shift $H_0$ furthest from the
\lcdm{} value, with preferred ranges reaching over $\sim70\,\kmsmpc$ and a tension reduced to the $\sim 3\sigma$ level,
while the varying electron-mass and WZDR models reaches intermediate values ($\sim 4\sigma$ and $\sim 4.5\sigma$ residual tension respectively). All other models remain close to the \lcdm{}
inference. Once the \mb{} calibration is included, the two statistical
frameworks grade the five groups as summarized below.

\begin{table}[!t]
\caption{\label{tab:selection}Baseline performance of all 13 contenders,
ordered by Groups E, M, R, and L. Tensions are quoted in Gaussian-equivalent
standard deviations. A model qualifies if either
$-\Delta{\rm AIC}>10$ or $\ln{\rm BF}>3$.
Dashes mark model-comparison metrics that are not defined for the
reionization-agnostic case since it involves a different dataset. }
\squeezetable
\begin{ruledtabular}
\begin{tabular}{lccccc}
Model & $-\Delta{\rm AIC}$ & $\ln{\rm BF}$ &
$\Delta_{\rm DMAP}$ & $\Delta_{\rm shift}^{M_B}$ & Status\\
\midrule \arrayrulecolor[HTML]{CCCCCC}
EDE           & 23.40 & 10.51 & 2.51 & 3.0 & Qualified\\
Cold NEDE     & 20.02 &  3.03 & 3.06 & 2.7 & Qualified\\
RnR           & 19.74 &  7.71 & 3.05 & 3.6 & Qualified\\
EMG           & 19.64 &  6.41 & 2.72 & 2.9 & Qualified\\
\midrule
Varying $m_e$ & 12.58 &  3.53 & 4.25 & 4.0 & Qualified\\
4param. Mod.Rec.  &  9.78 & -0.17 & 4.72 & 5.4 & Eliminated\\
\midrule
WZDR          &  7.06 &  1.77 & 4.45 & 4.5 & Eliminated\\
$\Delta N_{\rm eff}$ & 3.18 & -0.09 & 5.09 & 5.1 & Eliminated\\
DRMD          &  0.86 & -6.44 & 4.90 & 5.1 & Eliminated\\
SIDR          & -0.69 & -1.42 & 5.26 & 5.0 & Eliminated\\
\midrule
Thaw. Gravity &  2.34 & -6.12 & 5.85& 5.3 & Eliminated\\
\midrule
$\Lambda_s$CDM&  3.60 &  0.92 & 5.15 & 5.5 & Eliminated\\
iDM--DE       & -2.67 & -3.90 & 5.45 & 5.4 & Eliminated\\
Agnos. reion. & ---   & ---   & 5.41 & 5.5 & Eliminated\\
\arrayrulecolor{black}
\end{tabular}
\end{ruledtabular}
\end{table}

All four Group E models cross the selection thresholds and reduce the residual
tension most efficiently. Their internal ordering depends mildly on the
statistic: AIC and $\Delta_{\rm DMAP}$ depend on best fits, whereas the evidence
and $\Delta_{\rm shift}$ respond to posterior volume. This reshuffling does not
alter the physical conclusion that a localized non-radiative contribution near
matter--radiation equality is the most effective mechanism in the baseline
comparison.

  Varying the electron mass is the only contender outside Group E to qualify for
the knockout round of paper II, although it leaves a larger residual discrepancy. The flexible
four-parameter deformation of the ionization history improves the best fit, but
its added freedom is penalized by both AIC and the Bayesian evidence, and it
does not cross either selection threshold. The CMB permits some
recombination freedom, but the joint CMB+BAO+SN geometry limits its ability to
follow the local calibration.

None of the radiation or late-time models crosses the selection thresholds.
WZDR nevertheless gives the best performance in Group R.  DRMD is strongly disfavored by the Bayesian
evidence, while SIDR and free-streaming $\Delta N_{\rm eff}$ do not decrease the tension significantly compared to \lcdm{}. Group L performs worst overall, with tension metrics similar to \lcdm{}. Finally, discarding low-multipole
polarization information changes the dataset itself and is therefore not
assigned an AIC or evidence ranking; it also does not reduce the calibration
tension.

The agreement between the two statistical approaches is itself informative:
AIC and Bayesian evidence penalize complexity differently, while
$\Delta_{\rm DMAP}$ depends on best fits whereas $\Delta_{\rm shift}$ depends on
posterior volume. Nevertheless, both statistical frameworks select Group E, identify the varying
electron mass as an intermediate case, and reject the claim that the tested
radiation-only or late-time mechanisms solve the tension for this baseline
dataset. The residual differences are of degree rather than kind: RnR has a
frequentist tension of $3.0\sigma$ but a Bayesian \mb{} shift of $3.6\sigma$,
and the broad modified-recombination parameter space improves the AIC
more than the evidence.

When comparing with previous analyses in which the summed neutrino mass was fixed, we find that fluctuating $\Sigma m_\nu$ in each model has no significant impact on the Hubble tension (in agreement with Ref.~\cite{Reeves:2022aoi}). We also find that the neutrino mass constraints are roughly as tight in the extended models as in the $\Lambda$CDM model ($\lesssim10\%$ difference), with the notable exception of the varying electron mass model, which relaxes this constraint by a factor of two (see Paper II).

 The baseline competition gives a concise empirical target for model building.
Successful models must reduce the sound horizon while preserving the detailed
high-multipole CMB spectra and the low-redshift distance relation fixed jointly
by DESI BAO and Pantheon+ SN. A transient scalar-field dark energy component near matter--radiation equality currently satisfies these requirements better than the other mechanisms tested here. Models with additional radiation are limited by
damping-tail and phase-shift signatures, while post-recombination changes have
too little freedom once BAO and SN are included. Modified recombination can
shift the relevant scales, but its allowed parameter space is narrowed by BAO and SN data that disfavor the smaller fractional matter density $\Omega_{\rm m}$ typical in these models.

Relative to the original $H_0$ Olympics \cite{Schoneberg:2021qvd}, the
radiation and recombination contenders have lost ground. We show in
Paper~II \cite{H0WorldCup:companion} that the strengthened radiation
constraints are driven by ACT data: removing ACT, or relaxing the assumed shape
of the primordial power spectrum through running of the spectral
index~\cite{Kosowsky:1995aa,Garny:2026gcs}, loosens them substantially, with
residual tensions dropping to the $3$--$3.5\sigma$ level. Likewise, allowing
spatial curvature improves the varying-electron-mass model, reducing
$\Delta_{\rm DMAP}$ from $4.3\sigma$ to $3.5\sigma$, though the benefit is
weaker than in the original Olympics---where varying $m_e$ with curvature was
the best-performing combination---because the combined CMB data now constrain
$\Omega_k$ much more tightly \cite{H0WorldCup:companion}.

Nevertheless, the outcome of the group stage is instructive. Even against a CMB+BAO+SN
combination far more constraining than that available to the original
Olympics, a coherent class of early-energy models crosses both frequentist and
Bayesian selection thresholds, shifts the inferred $H_0$ to
$\sim 70\,\kmsmpc$, and cuts the residual calibration tension from more than
$5\sigma$ to the $2.5$--$3\sigma$ level. Admittedly, no models reduce tension to a fully satisfactory level, nor are favored over $\Lambda$CDM in the absence of local $H_0$ information. However, these models are phenomenological proxies for certain classes of physical mechanisms. In fact, proposals to go further already exist: multi-field
realizations of EDE, for instance, significantly relax the tight constraints
imposed by Planck NPIPE data on the single-field model
\cite{Bella:2026mfe}. 

Modified recombination offers a parallel direction. A consistent quantum field theory permitting sufficient electron-mass variation while satisfying fifth-force bounds remains challenging \cite{Schoneberg:2024ynd,Baryakhtar:2024rky,Smith:2025grk}. This toy model is mostly useful to capture the key degeneracy that can weaken constraints on \(H_0\) in modified-recombination scenarios. Alternative
model-independent reconstructions admit broader ionization histories and can
favor earlier, skewed recombination once late-time data are included
\cite{Lynch:2024gmp,Lynch:2024hzh}, while primordial magnetic fields furnish a
physical realization supported by magnetohydrodynamic simulations and
Lyman-$\alpha$ radiative transfer \cite{Jedamzik:2025pmf}. This motivates
testing recombination histories beyond the proxies considered here. 

Moreover, the models considered in this contest also have potentially testable consequences for large-scale structures and Big Bang nucleosynthesis, as explored in Paper II. Beyond further scrutiny of the astrophysical assumptions and systematic uncertainties affecting \(H_0\) measurements, it will be important to investigate fundamental theories yielding the phenomenology we study here, identify complementary signatures—such as cosmic birefringence, fifth forces, or new particles accessible at accelerators— and perhaps  consider even more radical departures from \(\Lambda\)CDM altogether.

\textbf{\textit{Acknowledgments.}} We thank Lennart Balkenhol, Gia Dvali, Ali Rida Khalife, Florian Niederman, Martin Sloth, and Gen Ye for useful discussions. We thank Tristan L. Smith for initial contributions as well as useful discussions throughout. V.P.\ acknowledges the European Union's Horizon Europe research and innovation programme under the Marie Sk\l odowska-Curie Staff Exchange grant agreement no.\ 101086085 -- ASYMMETRY. This work received funding support from the European Research Council (ERC) under the European Union's HORIZON-ERC-2022 (grant agreement no.\ 101076865). This publication is based upon work from the COST Action CA21136 ``Addressing observational tensions in cosmology with systematics and fundamental physics'' (CosmoVerse), supported by COST (European Cooperation in Science and Technology). The results obtained in this paper were computed through resources from the Universe and Particles Laboratory of Montpellier (LUPM). We thank LUPM for providing the technical support, computing and storage facilities. N.S. acknowledges support from the Excellence Cluster ORIGINS which is funded by the Deutsche Forschungsgemeinschaft (DFG, German Research Foundation) under Germany’s Excellence Strategy - EXC-2094/2 - 390783311, as well as the funding through a Fraunhofer-Schwarzschild
Fellowship at the LMU. RKS
thanks the Alexander von Humboldt Foundation for their
support.
JL and MM acknowledge support from the DFG grant LE 3742/8-1. The authors gratefully acknowledge the computing time provided to them at the NHR Center NHR4CES at RWTH Aachen University (project number p0021792). AGF, FF, LM acknowledge the Open Physics Hub project (hosted by the University of Bologna) and INFN for granting access to their computational resources. FF, LM acknowledge partial financial support from the Progetti di Astrofisica Fondamentale INAF 2023, from INFN InDark initiative, from the contract ASI/INAF for the Euclid mission n. 2018-23-HH.0 and the ASI grant 2020-9-HH.0 (participation in LiteBIRD phase A).
The authors acknowledge the use of Anthropic's Claude AI, Google's Gemini 3.5, and OpenAI Codex as supplementary research-assistance tools for plotting, data analysis, and text editing. 
We also acknowledge the use of the Python packages \texttt{NumPy}, \texttt{SciPy}, \texttt{Matplotlib}, \texttt{GetDist},  \texttt{tensiometer}, \texttt{liquidcosmo}, \texttt{MCevidence}, \texttt{Prospect}, and \texttt{procoli}.
\bibliography{bibliography}
\end{document}